\documentstyle[12pt]{article}

\newcommand{\EQ}{\begin{equation}}
\newcommand{\EN}{\end{equation}}

\newcommand{\bear}{\begin{eqnarray}}
\newcommand{\ear}{\end{eqnarray}}

\begin{document}

\topmargin 0pt
\oddsidemargin 5mm
\newcommand{\NP}[1]{Nucl.\ Phys.\ {\bf #1}}
\newcommand{\PL}[1]{Phys.\ Lett.\ {\bf #1}}
\newcommand{\NC}[1]{Nuovo Cimento {\bf #1}}
\newcommand{\CMP}[1]{Comm.\ Math.\ Phys.\ {\bf #1}}
\newcommand{\PR}[1]{Phys.\ Rev.\ {\bf #1}}
\newcommand{\PRL}[1]{Phys.\ Rev.\ Lett.\ {\bf #1}}
\newcommand{\MPL}[1]{Mod.\ Phys.\ Lett.\ {\bf #1}}
\newcommand{\JETP}[1]{Sov.\ Phys.\ JETP {\bf #1}}
\newcommand{\TMP}[1]{Teor.\ Mat.\ Fiz.\ {\bf #1}}

\renewcommand{\thefootnote}{\fnsymbol{footnote}}

\newpage
\setcounter{page}{0}
\begin{titlepage}
\begin{flushright}
UFSCARF-TH-94-12
\end{flushright}
\vspace{0.5cm}
\begin{center}
{\large  Exact solution of the simplest super-orthosymplectic invariant
magnet}\\
\vspace{1cm}
\vspace{1cm}
{\large M.J.  Martins  } \\
\vspace{1cm}
{\em Universidade Federal de S\~ao Carlos\\
Departamento de F\'isica \\
C.P. 676, 13560~~S\~ao Carlos, Brasil}\\
\end{center}
\vspace{1.2cm}

\begin{abstract}
We present the exact solution of the $Osp(1|2)$ invariant magnet by the Bethe
ansatz approach.
The associated Bethe ansatz equation exhibit a new feature by  presenting an
explicit and distinct
phase behaviour in even and odd sectors of the theory. The ground state, the
low-lying excitations and
the critical properties are discussed by exploiting the Bethe ansatz solution.
\end{abstract}
\vspace{.2cm}
\centerline{PACS numbers: 05.50+q, 64.60.Cn, 75.10.Hk, 75.10.Jm}
\vspace{.2cm}
\centerline{September 1994}
\end{titlepage}

\renewcommand{\thefootnote}{\arabic{footnote}}
\setcounter{footnote}{0}

\newpage
The concept of integrability based on the Yang-Baxter relation can be extended
in order to include
in its algebraic structure variables possessing both commuting and
anticommuting rules of
permutation. This makes it possible to construct integrable lattice models in
which their bond
configurations admit bosonic and fermionic degrees of freedom \cite{KS}. This
notion is then
systematized by relating the Boltzmann weights of the lattice model to the
$R$-matrix solutions of
a graded version \cite{KS} of the Yang-Baxter equation. This generalization is
necessary in order to
take into account the Grassmann parities coming from the interchange of
fermions of the theory. Moreover,
for a typical system consisting of $n$ species of bosons and $m$ species of
fermions, their solutions
are believed to be found as invariants under the $Sl(n|m)$ and $Osp(n|2m)$
superalgebras \cite{BS}.

The exact solution of the $Sl(n|m)$ invariant systems has been studied in many
different contexts
in the literature. For instance, the $Sl(n|m)$ property  is resembled much on
the Perk-Schultz
vertex model \cite{PK,DE} which possesses an extra (discrete) parameter playing
the role of the $Sl(n|m)$
Grasmann parities. More recently, these solutions have also been examined in
strongly correlated
eletronic systems, being the one dimensional supersymmetric  $t-J$ model the
typical example
\cite{TJ,TJ1,TJ2}. However, for the $Osp(n|2m)$ series, not much is known
concerning its
Bethe ansatz solutions and critical properties. As far as we know, the only
exception is the $U_qOsp(2|2)$ \cite{ITO} model which has been
recently solved in the context of $N=2$ supersymmetric lattice models
\cite{MS}.

The purpose of this letter is to exactly solve the simplest system in the
$Osp(n|2m)$ family, namely
the case of one boson ($n=1$) and two fermions ($m=1$). As we shall see,
a new feature appears in the
Bethe ansatz equations, making possible an explicit distinction between the
even and odd sectors of the
model. This new property has important consequences for the finite-size
behaviour which governs the
class of universality of this theory.

We begin by introducing the isotropic $Osp(1|2)$ magnet. Its Hamiltonian
defined on a
lattice of $L$ sites (assuming periodic boundary conditions)  can be
written in the following compact form
\EQ
H= -\sum_{i=1}^{L} [ P_{i,i+1} +\frac{2}{3} E_{i,i+1} ]
\EN
where $(P_{i,i+1})_{ab}^{cd} = (-1)^{p(a)p(b)} \delta_{a,d} \delta_{b,c}$ is
the graded permutation
operator and $(E_{i,i+1})_{ab}^{cd} =\alpha_{ab} \alpha_{cd}^{st}$ is the
$Osp(1|2)$ Temperely-Lieb
invariant operator \cite{MP}. The variables $p(a)$ are the Grasmann parities,
namely $p(1)=0$ (boson)
and $p(2)=p(3)=1$ (fermions). The symbol $\alpha^{st}$ denotes the
supertranspose operation on a $3X3$
matrix $\alpha$ given by
\EQ
\alpha=\left( \begin{array}{ccc}
	1 &   0 & 0 \\
	0 &  0& 1 \\
	0 & -1 & 0 \\
	\end{array}
	\right)
\EN

The Bethe ansatz solution of (1) is as follows. The first step is to write the
corresponding
vertex operator of the lattice model leading to the Hamiltonian (1). After an
appropriate
canonical transformation, it is possible to show that this weight can be
expressed in terms of
the generators $S^z,S^+,S^-$ of a spin-1 algebra. Its explicit form is
\EQ
R(\lambda)= \left( \begin{array}{ccc}
          \lambda I +f(\lambda) S^{z} -g(\lambda)(S^{z})^{2} &
\frac{1}{\sqrt{2}}[S^- S^z
-2\lambda h(\lambda) S^{z} S^{-}] & f(\lambda) (S^-)^2 \\
           \frac{1}{\sqrt{2}}[S^z S^+
-2\lambda h(\lambda) S^{+} S^{z}] & [\lambda -3 h(\lambda)]I +3 h(\lambda)
(S^z)^2 &
           -\frac{1}{\sqrt{2}}[S^z S^-
+2\lambda h(\lambda) S^{-} S^{z}]  \\
 f(\lambda) (S^+)^2 &
           -\frac{1}{\sqrt{2}}[S^+ S^z
+2\lambda h(\lambda) S^{z} S^{+}]  &
          \lambda I -f(\lambda) S^{z} -g(\lambda)(S^{z})^{2} \\
\end{array}
\right)
\EN
where $I$ is the 3X3 identity matrix and
functions $f(\lambda)$, $g(\lambda)$ and $h(\lambda)$ are given by
\EQ
f(\lambda)=\frac{(2\lambda-3/2)}{2(\lambda-3/2)};~~
h(\lambda)=\frac{1}{2(\lambda-3/2)};~~g(\lambda)=
2\lambda +\frac{3}{2}h(\lambda)
\EN

It is important to notice that the reference state preserving the triangularity
property of this $R$-matrix is still the usual ferromagnetic pseudo-vaccum.
Moreover, we recall that
its structure resembles much the one that appears in the Izergin-Korepin (IK)
model \cite{IK,RE}. Indeed we
are able to show that the Hamiltonian (1) can be obtained as a certain branch
limit of the IK system
through a delicate but rigorous canonical transformation. The calculations
leading to this
result are rather cumbersome and will be presented elsewhere \cite{MA}. Here we
can use this information
,by adapting  the algebraic and(or) the analytic Bethe ansatz method used in
the IK model \cite{RE},
in order to solve
Hamiltonian (1). Due to a $U(1)$-conserved charge the $Osp(1|2)$ Hamiltonian
can be block
diagonalized into disjoint sectors labelled by the total magnetization index,
namely
$r=\sum_{i=1}^{L} S_i^z$. In a given sector $r$, for periodic boundary
condition, the eigenstates are
parametrized by a set of complex numbers $\lambda_j$ satisfying the following
Bethe ansatz equation
\EQ
{\left( \frac{\lambda_j -i/2}{\lambda_j +i/2} \right)}^{L}= - (-1)^r
\prod_{k=1}^{L-r}
\left(\frac{\lambda_j -\lambda_k -i}{\lambda_j-\lambda_k +i} \right)
\left(\frac{\lambda_j -\lambda_k +i/2}{\lambda_j-\lambda_k -i/2} \right)
\EN
and the eigenenergies are given by
\EQ
E_r(L) =-\sum_{j=1}^{L-r} \frac{1}{\lambda_j^2 +1/4} + L
\EN

The important novelty present in the Bethe ansatz equation is its dependence
on a phase factor $(-1)^r$ even
for a system with periodic boundary conditions. Hence, even and odd sectors of
the theory are
distinguished by the signs of their ``bare'' phase-shift (right hand side of
(4) ). This may
be interpreted
as an explicit separation of the bosonic and fermionic degrees of freedom
present in the system. In fact,
we have verified this ``sector separation'' through
an exact diagonalization of the Hamiltonian (1) for small
values of $L$. We also remark that although a phase factor does not affect the
basic properties in the
thermodynamic limit ($L \rightarrow \infty$), they do change the finite-size
behaviour. Therefore, this
phase factor plays an important role in order to characterize the correct
critical behaviour of the model.

Before turning to the study of the finite size properties of (4) and (5), we
first compute some
useful quantities in the thermodynamic limit. The ground state in each sector
$r$ is parametrized by a set
of real numbers $\lambda_j$. Taking  the logarithm of equation (4) and
correctly collecting  the phase
factors, we rewrite the Bethe ansatz equations as
\EQ
L \psi_{1/2}(\lambda_j) =2 \pi Q_j + \sum_{k=1,k \neq j}^{L-r}[
\psi_{1}(\lambda_j-\lambda_k) -
\psi_{1/2}(\lambda_j-\lambda_k)]
\EN
where $\psi_a(x)= 2\arctan(x/a)$ and $Q_j$ are integer or semi-integer series
of number
\EQ
Q_j= -\frac{[L-r-1]}{2} +j-1,~~ j=1,2, \cdots, L-r
\EN

For large $L$, the roots $\lambda_j$ are densely packed into its density
distribution,
$\sigma_L(\lambda_j) \cong 1/L(\lambda_{j+1}-\lambda_j)$. Strictly in the $L
\rightarrow \infty $
limit this system can be solved by elementary Fourier techniques and we find
that
\EQ
\sigma(\lambda) =\frac{2}{\sqrt{3}} \frac{\cosh(2 \pi \lambda/3)}{\cosh(4 \pi
\lambda/3) +1/2}
\EN

The ground state energy (per site) can be calculated by using equations (6) and
(9) after replacing the
sum by an integral. The final result is
\EQ
e_{\infty} = \lim_{ L \rightarrow \infty} E_0(L)/L = -\frac{4 \pi \sqrt{3}}{9}
+1 \cong -1.4184
\EN

In addition, we have verified that the low-lying excitations over the ground
state are gapless.
As usual, the excitations are built up by inserting vacancies on the density
distribution of
$\lambda_j$. The simplest excitation is the triplet state made of two real
``holes'' $\lambda_1$
and $\lambda_2$. Their contribution to the total momenta is calculated to be
\EQ
p(\lambda_1,\lambda_2) = \pi -\sum_{i=1}^{2} 2 \arctan[ \sinh(2\pi \lambda_i
/3)/ \cosh(\pi/6)]
\EN
which leads to the following low-momentum dispersion relation
\EQ
\epsilon(p) \sim v_s p
\EN
where $v_s =2 \pi/3$ is the sound velocity.

At this point we have obtained the most important quantities necessary for the
analysis of the
critical properties. The class of universality can be
determined by exploiting a set of important relations between the eigenspectrum
of
finite-lattice system \cite{CA}. For instance, the central charge $c$ is
related to the ground state
energies $E_0(L)$ by \cite{CA1}
\EQ
E_0(L)/L -e_{\infty} \cong -\frac{\pi c v_s}{6L^2}
\EN

In Table 1 we present our results for the central charge $c$. This result is
obtained by numerically solving the Bethe ansatz equation in the sector $r=0$.
The extrapoled results
predict a conformal anomaly $c=1$. The sectors $r \neq 0$ are responsible for
the low-lying
conformal dimensions. In this case , however, a numerical study up to lattice
size $L \sim 64$
presents a poor convergence rate. This is due to the appearance of
logarithm corrections of order $O(1/\ln(L))$ in the scaling function.
Fortunately, since the roots $\lambda_j$ are real, we can use an analytical
method developed
by De Vega and Woynarovich \cite{DV} in order to estimate these critical
dimensions. The calculation
is quite technical \cite{MA} and here we only present the final results. We
stress, however,
that the phase factor of (5) is of fundamental importance in this computation.
The  critical dimension
$X_r$ associated with the lowest operator on sector $r$ has the following form
\EQ
X_r =\frac{r^2}{4}
\EN

Such dimension corresponds to a spin-wave excitation of index $r$ in a Gaussian
structure of
operators. In fact, more generally we expect to find the following dimensions
\EQ
X_{r,m} = \frac{r^2}{4} +m^2
\EN
where $m$ stands for the ``vortex '' excitations index. On the other hand, this
result can be
justified by using the following arguments of symmetry. For example, we  have
verifed that (
for finite lattices) the first excited state with zero momentum in the $r=0$
sector is degenerated
to the ground state in sector $r=2$. This means the identity $X_{0,1}=X_{2,0}$,
which is indeed verified in
formula (15). Evidently, the same argument can be repeated for many other
values of $r$ and $m$,
confirming the conjecture (15). We recall that this is the manifestation of the
$Osp(1|2)$ symmetry
encoded in the important phase factor $(-1)^r$.

In summary, we have given a Bethe ansatz solution to the eigenvalues of the
$Osp(1|2)$ magnet.
The novelty is the presence of a phase factor depending on the sector $r$,
which becomes crucial in order
to determine  its critical properties. We believe that our results extends the
form of Bethe
ansatz solution appearing in integrable lattice models. Hopefully, our results
will be
of relevance in the discussion of the Bethe ansatz completeness and  in the
thermodynamic properties.

\section*{Acknowledgements}
It is a pleasure to thank M. Malvezzi for his help with numerical checks and
F.C. Alcaraz for
innumerous discussions. This work is
supported by CNPq (Brazilian agency).
\vspace{0.4cm}\\

\vspace{0.5cm}

\centerline{ \bf Table Captions}
\vspace{0.5cm}
Table 1. The estimative of the conformal anomaly of equation (13)

\begin{center}
{\bf Table 1}\\
\vspace{0.5cm}
\begin{tabular}{|l|l|} \hline
L & $ -6L^2[E_0(L)/L-e_{\infty}]/\pi v_s$ \\ \hline \hline
16 & 1.0029 729 \\ \hline
24 & 1.0022 324  \\ \hline
32 &  1.0018 406 \\ \hline
40 & 1.0015 955 \\ \hline
48 & 1.0014 261 \\ \hline
56 & 1.0013 010 \\ \hline
64 & 1.0012 043 \\ \hline
Extrapolated & 1.0001 (1) \\ \hline
\end{tabular}
\end{center}

\end{document}